\documentclass[conference]{IEEEtran}
\IEEEoverridecommandlockouts

\usepackage{cite}
\usepackage{amsmath,amssymb,amsfonts}
\usepackage{algorithmic}
\usepackage{graphicx}
\usepackage{booktabs}
\usepackage{textcomp}
\usepackage{xcolor}
\usepackage{pgfplots}
\pgfplotsset{compat=1.8}
\usepackage{tikz}
\usetikzlibrary{spy}
\usepackage{soul}
\usepackage{pgfplotstable}
\usepgfplotslibrary{statistics}
\usepackage{fancyhdr}
\pgfplotsset{grid style={dotted,gray}}
\usepackage{lipsum}
\def\BibTeX{{\rm B\kern-.05em{\sc i\kern-.025em b}\kern-.08em
    T\kern-.1667em\lower.7ex\hbox{E}\kern-.125emX}}
\begin{document}

\title{Learning Binary Autoencoder-Based Codes with Progressive Training\\
\thanks{This research has been supported by the Ministry of Science, Technological
Development and Innovation (Contract No. 451-03-137/2025-03/200156) and
the Faculty of Technical Sciences, University of Novi Sad through project
“Scientific and Artistic Research Work of Researchers in Teaching and
Associate Positions at the Faculty of Technical Sciences, University of Novi Sad
2025” (No. 01-50/295).}
}

\author{\IEEEauthorblockN{Vukan Ninkovic\IEEEauthorrefmark{1}\IEEEauthorrefmark{2}, Dejan Vukobratovic\IEEEauthorrefmark{2}
\vspace{1mm}
\IEEEauthorblockA{
\IEEEauthorblockA{\IEEEauthorrefmark{1}The Institute for Artificial Intelligence Research and Development of Serbia, Serbia
}
\IEEEauthorblockA{\IEEEauthorrefmark{2}Faculty of Technical Sciences, University of Novi Sad, Serbia}
}}}

\maketitle

\begin{abstract}
Error-correcting codes play a central role in digital communication, ensuring that transmitted information can be accurately reconstructed despite channel impairments. Recently, autoencoder (AE)-based approaches have gained attention for the end-to-end design of communication systems, offering a data-driven alternative to conventional coding schemes. 
However, enforcing binary codewords within differentiable AE architectures remains difficult, as discretization breaks gradient flow and often leads to unstable convergence. 
To overcome this limitation, a simplified two-stage training procedure is proposed, consisting of a continuous pretraining phase followed by direct binarization and fine-tuning without gradient approximation techniques. 
For the $(7,4)$ block configuration over a binary symmetric channel (BSC), the learned encoder–decoder pair learns a rotated version (coset code) of the optimal Hamming code, naturally recovering its linear and distance properties and thereby achieving the same block error rate (BLER) with maximum-likelihood (ML) decoding.
These results indicate that compact AE architectures can effectively learn structured, algebraically optimal binary codes through stable and straightforward training.
\end{abstract}


\section{Introduction}

Reliable digital communication systems represent information as sequences of binary symbols that must be efficiently encoded to ensure robustness against channel impairments \cite{lee_1994}. 
Classical error-correcting codes, such as the Hamming and BCH families, achieve optimal or near-optimal performance for short block lengths through carefully designed algebraic structures \cite{allum_2016}. However, these traditional codes depend on manually derived constructions and decoding algorithms that are often tailored to specific channel models or fixed code rates. 
With the growing influence of machine learning in physical-layer design, autoencoder (AE) architectures have emerged as a flexible data-driven alternative, capable of jointly learning the transmitter and receiver mappings in an end-to-end fashion directly from data \cite{OShea_2017}.

Despite their conceptual appeal, obtaining binary codeword representations from AEs remains challenging. Standard gradient-based optimization inherently favors continuous latent spaces, while discrete outputs introduce non-differentiability that disrupts gradient flow during training. This difficulty is well recognized in representation learning and hashing literature, where several methods have been proposed to mitigate it, including binary AEs~\cite{hashing_binary_AE}, neural discrete representation learning~\cite{van2017}, and discrete AEs for sequence models~\cite{kaiser2018}. These works consistently report that direct discretization of latent features leads to vanishing or unstable gradients, necessitating specialized relaxation or approximation techniques.

In the communications domain, a number of studies have explored the integration of binary constraints into learned channel coding frameworks. Notable examples include the turbo AE~\cite{turbo_AE}, trainable communication systems based on the binary neural network~\cite{che_2022}, and neural joint source–channel coding~\cite{neural_JSC}. These approaches demonstrate the feasibility of end-to-end trainable digital communication systems, yet they often rely on complex training procedures such as stochastic binarization \cite{hubara_2016}, gradient estimators \cite{fu_2006}, or straight--through estimation \cite{yin_2017}, which enable approximate gradient propagation during
training, but can complicate convergence and hinder practical deployment.

Beyond these works, several recent studies have extended the scope of deep learning-based code design toward alternative communication paradigms. In~\cite{cohen_2023}, the Advanced Encryption Standard (AES) cryptosystem was shown to function as an error-correcting mechanism, achieving reliability comparable to random linear codes through a simple padding–encryption composition. Similarly, an AE-based framework for code-domain non-orthogonal multiple access (CD-NOMA)~\cite{han_2022} jointly optimized resource mapping and codebook design, approaching the bit-error-rate (BER) performance of a single-user modulation system. Another related work, the Neural Belief Propagation AE (NBP-AE)~\cite{larue_2022}, investigated the joint learning of linear block codes and BP-like decoders, offering an interpretable and scalable design with favorable complexity–performance trade-offs. Collectively, these studies highlight the versatility of deep learning for coding theory but also reveal that achieving stable binary representations within differentiable training pipelines remains a major challenge.

Motivated by these challenges, this paper proposes a simplified AE-based framework for learning binary communication codes. The proposed approach adopts a two-phase training strategy: an initial continuous pretraining stage that ensures stable convergence, followed by a direct binarization and fine-tuning phase that yields valid binary codewords without relying on gradient approximation techniques.Through this design, we demonstrate that for a short block length $(7,4)$ configuration, the proposed AE-based system learns a rotated version (coset code) of the optimal Hamming code, naturally recovering its linear structure and distance spectrum. Consequently, the learned encoder–decoder pair achieves the same block error rate (BLER) performance as the Hamming$(7,4)$ code with ML decoding.

\section{Background \& System Overview}
\subsection{System Overview}
\label{sec:system_model}
This work addresses the reliable transmission of a discrete information message $m$ over a binary noisy communication channel. 
Let $\mathcal{M} = \{1,2,\ldots,M\}$ denote the set of possible source messages, where each message is represented by a sequence of $k$ information bits, so that $M = 2^{k}$. 
At the transmitter, the message $m \in \mathcal{M}$ is encoded into a binary codeword $\boldsymbol{x}$ of length $n$ through a mapping $f : \mathcal{M} \rightarrow \{-1, +1\}^{n}$,  
    $\boldsymbol{x} = f(m) = (x_1, x_2, \ldots, x_n)$.
The collection of all possible codewords forms the codebook
    $\mathcal{X} = \{ f(m) \mid m \in \mathcal{M} \} \subset \{-1, +1\}^{n},$
which inherently satisfies the total energy constraint $ \| \boldsymbol{x} \|_2^2 = n,$ $\forall \boldsymbol{x} \in \mathcal{X}.$ 
Accordingly, the code rate is given by
$    R = \frac{k}{n} \quad \text{[bits/channel use]}.$

The transmission medium is modeled as a discrete memoryless channel denoted by $\mathcal{C}(\cdot)$, 
which transforms the transmitted binary codeword $\boldsymbol{x}$ into the received sequence $\boldsymbol{y}$. 
For a binary symmetric channel (BSC)\footnote{Although this work focuses on the BSC, the proposed framework can be readily extended to other channel models such as the binary erasure channel (BEC) or additive noise channels.} with crossover probability $p$, each transmitted bit is independently flipped with probability $p$.
Equivalently, the channel operation can be expressed as:
\begin{equation}
    \boldsymbol{y} = \mathcal{C}(\boldsymbol{x}) = \boldsymbol{x} \odot \boldsymbol{z},
\end{equation}
where $\odot$ denotes the elementwise (Hadamard) product and 
$\boldsymbol{z} = (z_1, z_2, \ldots, z_n)$ is a random noise vector with independent components:
\begin{equation}
    z_i = 
    \begin{cases}
        +1, & \text{with probability } 1-p,\\
        -1, & \text{with probability } p.
    \end{cases}
\end{equation}
The received sequence $\boldsymbol{y} = (y_1, y_2, \ldots, y_n)$ thus belongs to the same binary alphabet as $\boldsymbol{x}$.

At the receiver, the noisy codeword $\boldsymbol{y} \in \{-1,+1\}^n$ is processed by a decoder mapping
    $g : \{-1,+1\}^{n} \rightarrow \mathcal{M}, \quad 
    \hat{m} = g(\boldsymbol{y}),$
 where $\hat{m}$ denotes the receiver's estimate of the transmitted message $m$. 
For a given codebook $\mathcal{X}$ and a channel realization, the decoder typically selects the message whose corresponding codeword is most likely to have generated the received vector. 
In the case of a BSC, this corresponds to choosing the codeword that minimizes the Hamming distance to $\boldsymbol{y}$:
\begin{equation}    \hat{m} = \arg \min_{m' \in \mathcal{M}} d_H\big(\boldsymbol{y}, f(m')\big),
\end{equation}
where $m' \in \mathcal{M}$ is a candidate message, $f(m')$ is its corresponding codeword, and $d_H(\cdot, \cdot)$ denotes the Hamming distance between two binary sequences.  
The overall goal of the communication system is to design the encoder–decoder pair $(f, g)$ that, for a certain channel model, minimizes the average message error probability:
\begin{equation}
\label{eq: avg_prob}
    P_e = \frac{1}{M} \sum_{m \in \mathcal{M}} \Pr\big(\hat{m} \neq m \,\big|\, m\big).
\end{equation}

\subsection{Autoencoder--Based Communication System}
\label{AE_system}

\begin{figure}
    \centering
    \includegraphics[width=1\linewidth]{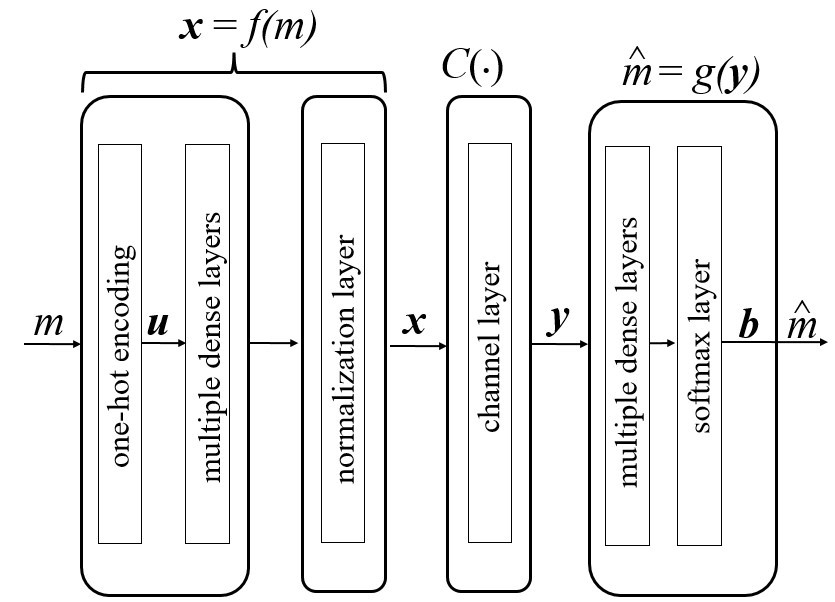}
    \caption{AE--based communication system \cite{OShea_2017}.}
    \label{fig:AE}
\end{figure}

The conventional digital communication system described above can be reformulated as an AE \cite{OShea_2017}, where the transmitter and receiver correspond to the encoder and decoder networks, respectively, as shown in Fig. \ref{fig:AE}. In this representation, the encoding function $f(\cdot)$ and decoding function $g(\cdot)$ are both implemented as neural networks and jointly optimized in an end-to-end manner to minimize the overall message reconstruction error across the noisy channel.

At the transmitter side, each source message $m \in \mathcal{M}$ is first represented as a one-hot vector $ \boldsymbol{u} = (u_1, u_2, \ldots, u_M) \in \{0,1\}^{M},$
where $u_i = 1$ if $i = m$ and $u_i = 0$ otherwise. 
This vector serves as the input to the neural encoder, typically realized by a series of fully connected (FC) layers, whose goal is to learn a latent representation that corresponds to the most suitable codeword $\boldsymbol{x} = f(m) \in \mathcal{X}$ for reliable transmission over the channel (Fig. \ref{fig:AE}).

In most existing AE-based communication systems, 
the output of the encoder represents a continuous-valued vector. 
This design choice preserves differentiability of the end-to-end learning process 
and facilitates gradient-based optimization. 
However, it results in continuous codewords that differ from practical digital communication systems, 
where transmitted signals are inherently binary. 
Enforcing discrete codeword representations within the neural encoder 
is known to be challenging, as the discretization process is non-differentiable 
and may hinder gradient-based optimization \cite{van2017}. 

The effect of the communication channel is typically modeled as a 
non-trainable stochastic transformation $\mathcal{C}(\cdot)$ (Section \ref{sec:system_model}), that maps the transmitted codeword 
$\boldsymbol{x}$ to the received vector $\boldsymbol{y}$, i.e., $\boldsymbol{y} = \mathcal{C}(\boldsymbol{x}).$ The specific channel model (e.g., BSC or additive noise channel) defines the statistical relationship between $\boldsymbol{x}$ and $\boldsymbol{y}$ 
and is incorporated into the training loop as an independent, non-trainable layer (Fig. \ref{fig:AE}).

At the receiver, the decoding function is realized by a neural network that mirrors the structure of the transmitter’s encoder. Given the received vector $\boldsymbol{y}$, the decoder maps it to an estimate of the transmitted message as $\hat{m} = g(\boldsymbol{y}).$ The last layer of the decoder employs a softmax activation function, as shown in Fig. \ref{fig:AE}, producing an $M$-dimensional output vector $\boldsymbol{b} = (b_1, b_2, \ldots, b_M),$ where each element $b_i$ denotes the estimated probability that the transmitted message corresponds to index $i$, 
satisfying $b_i \in [0,1]$ and $\sum_{i=1}^{M} b_i = 1$. 
The final decision $\hat{m}$ is obtained as $\hat{m} = \arg\max_{i} b_i.$

Since the average message error probability (Eq.~\eqref{eq: avg_prob}) is non-differentiable, 
the AE is trained using the categorical cross-entropy loss as a differentiable surrogate between  the target one-hot vector $\boldsymbol{u}$ and the predicted probability vector $\boldsymbol{b}$, defined as:
\begin{equation}
    \mathcal{L} = - \sum_{i=1}^{M} u_i \log b_i.
    \end{equation}
This enables end-to-end optimization of both the encoder and decoder parameters using gradient-based learning.

\section{Autoencoder-Based Digital Binary Codes}
\subsection{Binary Codeword Formation}\label{sec:binary_codeword}
A central challenge in AE-based digital communication design is to produce binary codewords that can be transmitted over a digital channel, while maintaining differentiability required for gradient-based optimization.
If the encoder is forced to generate discrete $\{-1, +1\}$ outputs from the start of training, the resulting non-differentiability often disrupts gradient flow and hinders convergence \cite{turbo_AE}.

To overcome this, the proposed method introduces a
two-stage learning strategy consisting of:
\begin{enumerate}
    \item \textbf{Continuous pretraining phase}, during which the encoder--decoder pair is trained 
    using continuous-valued latent representations (codewords) to ensure stable convergence and to
    capture the underlying code structure.
    \item \textbf{Binarization and fine-tuning phase}, where the continuous encoder outputs are projected onto the binary alphabet $\{-1,+1\}^{n}$ to obtain valid digital codewords. In this phase, training continues directly on the binary representations. The convergence stability achieved during the continuous pretraining phase ensures that the network can successfully adapt to the discrete codewords without gradient approximation techniques.
\end{enumerate}

Let $\tilde{\boldsymbol{x}} = f(m) \in \mathbb{R}^{n}$ denote the continuous encoder output for message $m \in \mathcal{M}$ during pretraining. After the pretraining phase converges, the encoder output is binarized as
    $\boldsymbol{x} = \operatorname{sign}(\tilde{\boldsymbol{x}}),$
yielding the final binary codeword $\boldsymbol{x} \in \{-1,+1\}^{n}$. The binary codewords are transmitted over the noisy channel $\mathcal{C}(\cdot)$, producing the received vector $\boldsymbol{y} = \mathcal{C}(\boldsymbol{x})$. The decoder $g(\cdot)$ then estimates the transmitted message from $\boldsymbol{y}$ via an softmax output with dimension $M$ (Section \ref{AE_system}).

This procedure preserves the end-to-end nature of the AE framework, while ensuring that the final learned codebook:
\begin{equation}
    \mathcal{X} = \{ f(m) \mid m \in \mathcal{M} \} \subset \{-1,+1\}^{n}
\end{equation}
consists of discrete, energy-constrained codewords suitable for digital transmission, as defined in Section~\ref{sec:system_model}. Details on the network architecture and training procedure are provided in Section~\ref{sec:architecture}.

\subsection{Network Architecture \& Training Procedure}
\label{sec:architecture}
The proposed AE-based digital communication system consists of two fully connected neural networks that jointly realize the encoder and decoder functions. Both networks are designed with a simple feed-forward structure, where the intermediate layers are linear, i.e., they do not employ any nonlinear activation functions. This design choice was empirically found to provide more stable convergence and improved performance in the considered digital communication setup (as will be discussed in Section IV-A). The absence of intermediate nonlinearities allows the encoder to form linearly separable latent representations, reducing the risk of gradient saturation and making the subsequent binarization step more robust. Moreover, since both the input (one-hot message representation) and the encoder output (binary codeword) lie in well-structured discrete spaces, introducing additional nonlinearities between layers was found to yield no significant advantage.

At the transmitter, the encoder network $f(\cdot)$ consists of two fully connected layers. 
The input message $m \in \mathcal{M}$ is encoded by a one-hot vector $\boldsymbol{u} \in \mathbb{R}^{M}$, where $M = 2^{k}$. 
The first layer projects $\boldsymbol{u}$ into a latent space of the same dimension, while the second layer compresses this representation into an $n$-dimensional continuous vector:
$\tilde{\boldsymbol{x}} = f(m) \in \mathbb{R}^{n}.$
Batch normalization is applied to normalize the power of $\tilde{\boldsymbol{x}}$, followed by a $\tanh(\cdot)$ activation that constrains the output values to the interval $[-1, 1]$. 
This step enables a smooth transition toward the binary alphabet $\{-1, +1\}$, which is later achieved through the binarization procedure described in Section~\ref{sec:binary_codeword}.

At the receiver, the decoder network $g(\cdot)$ mirrors the structure of the encoder. 
The received vector $\boldsymbol{y}$ is first expanded to an intermediate representation of dimension $M = 2^{k}$ and then transformed into an $M$-dimensional output vector. 
The final layer applies a softmax activation function, producing the probability vector $\boldsymbol{b}$ over all possible transmitted messages. 
The estimated message is then obtained as $\hat{m} = \arg\max_i b_i$ (as defined in Section~\ref{AE_system}).

During the first 95 epochs, the encoder produces continuous-valued outputs, which facilitates stable gradient-based optimization and allows the network to learn the underlying code structure. 
After this pretraining stage, the encoder outputs are binarized according to
$\boldsymbol{x} = \operatorname{sign}(\tilde{\boldsymbol{x}}),$
and the training continues for the remaining epochs directly on the binary representations, without using gradient approximation techniques. 
This simple transition from continuous to binary training proved effective, yielding stable convergence and well-separated codewords suitable for digital transmission.

For each mini-batch, a random binary mask $\boldsymbol{m} \in \{-1,+1\}^{n}$ is generated to simulate BSC channel variations and improve robustness. 
Each bit of the mask is flipped with probability $p$, which is a hyperparameter sampled uniformly from the range $[0.06, 0.1]$, and the mask is applied elementwise to the encoder output prior to decoding. The training dataset consists of $100{,}000$ randomly generated message samples, while the test set includes $1{,}000{,}000$ samples. 
In both cases, the messages are drawn uniformly from the message set $\mathcal{M}$. 
Training is conducted for $150$ epochs with a mini-batch size of $10$, and the overall pipeline is optimized using the Adam optimizer~\cite{adam} with a learning rate of $9\times10^{-4}$.

\section{Performance Evaluation}
\subsection{BLER Performance over the BSC}
The performance of the proposed binary AE-based communication system is evaluated in terms of BLER over a BSC with crossover probability $p$. All simulations are carried out for block length $n = 7$ and message length $k = 4$, corresponding to a code rate of $R = 4/7$. In this setting, the classical Hamming$(7,4)$ code with ML decoding is known to be an optimal linear block code. For this reason, it is adopted as the primary reference benchmark throughout the performance evaluation.

\begin{figure}[!t]
	\begin{tikzpicture}[spy using outlines=
{rectangle, red, magnification=3.2, line width=2, connect spies}]
  	\begin{semilogyaxis}[width=1\columnwidth, height=7cm, 
	legend style={at={(0.77,0.4)}, anchor= north,font=\scriptsize, legend style={nodes={scale=1.1, transform shape}}},
   	legend cell align={left},
	legend columns=1,   	 
   	x tick label style={/pgf/number format/.cd,fixed,
   	 precision=0.01, /tikz/.cd},
   	y tick label style={/pgf/number format/.cd,fixed, precision=1, /tikz/.cd},
   	xlabel={Crossover probability $p$},
   	ylabel={BLER},
   	label style={font=\footnotesize},
   	grid=both,   
   	xmin =0.01, xmax = 0.1,
   	ymin=0.002, ymax=0.17,
   	line width=0.85pt,
   	xtick={0.01, 0.02, 0.03, 0.04, 0.05, 0.06, 0.07, 0.08, 0.09, 0.1},
   	xticklabels={$0.01$, $0.02$, $0.03$, $0.04$, $0.05$, $0.06$, $0.07$, $0.08$, $0.09$, $0.1$},   	
   	tick label style={font=\footnotesize},]
    \addplot[blue, mark=x] 
   	table [x={x}, y={y}] {./figs./hamming_conv.txt}; 
   	\addlegendentry{$f_{\text{Hamming}}$ - $g_{\text{ML}}$} 
   	
    \addplot[red, mark=*] 
   	table [x={x}, y={y}] {./figs./hamm_ae.txt}; 
   	\addlegendentry{$f_{\text{Hamming}}$ - $g_{\text{AE}}$}
    
        \addplot[black, mark=square] 
   	table [x={x}, y={y}] {./figs./ae_ml.txt}; 
   	\addlegendentry{$f_{\text{AE}}$ - $g_{\text{ML}}$}
    
       	\addplot[green, mark=diamond] 
   	table [x={x}, y={y}] {./figs./ae_pure.txt}; 
   	\addlegendentry{$f_{\text{AE}}$ - $g_{\text{AE}}$}

    \coordinate (spypoint) at (axis cs:0.03,0.018);
\coordinate (spyviewer) at (axis cs:0.03, 0.09);
\spy[width=1.2cm,height=1.2cm] on (spypoint) in node [fill=white] at (spyviewer);
  	\end{semilogyaxis}
	\end{tikzpicture}
	\vspace*{-5mm}
	\caption{BLER performances versus crossover probability $p$ for four different scenarios.}
	\label{fig:bler}
\end{figure}
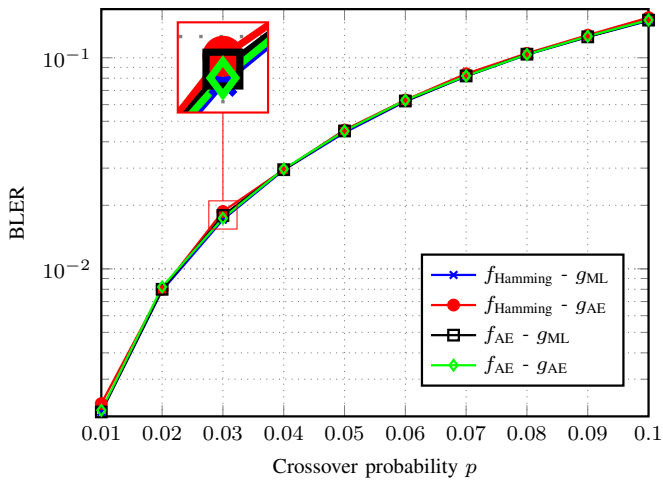

To assess both the learned codebook and decoding capability, the following configurations are compared:  
(i) the classical Hamming$(7,4)$ code with ML decoding
(ii) the classical Hamming code decoded using the trained neural decoder.  
(iii) the learned binary codebook combined with an ML decoder; and  
(iv) the proposed AE with learned encoder and decoder.    
This comparison allows us to separately examine (a) the quality of the learned codebook and (b) the generalization ability of the neural decoder to structured algebraic codes.

Fig.~\ref{fig:bler} illustrates the BLER as a function of the channel crossover probability $p$. The proposed AE achieves almost the same performance as the Hamming$(7,4)$ code with optimal ML decoding across the entire range of tested channel conditions. This demonstrates that the learned encoder–decoder pair is able to discover a codebook and decoding rule that are effectively equivalent to the optimal solution for this block length and rate. 

In addition, when the learned binary codebook is decoded using an ML decoder, its performance remains identical to that of the classical Hamming code. Conversely, when the trained neural decoder is applied to Hamming codewords, it achieves the same error performance as the ML decoder. These results indicate that the AE not only learns a near-optimal codebook but also generalizes to decode algebraic codes, suggesting that it implicitly captures the ML decision rule.

\subsection{Codebook Analysis: Linearity and Distance Spectrum}
\label{sec:codebook_analysis}

Beyond BLER performance, we examine the structure of the learned binary codebook to determine whether it exhibits algebraic properties similar to the Hamming$(7,4)$ code. Let $\mathcal{X} = \{ \boldsymbol{x}_1, \boldsymbol{x}_2, \dots, \boldsymbol{x}_{2^{k}} \}$ denote the set of binary codewords obtained after the binarization phase. Since the AE is not constrained to learn a linear code, there is no inherent guarantee that $\mathcal{X}$ forms a linear subspace of $\{-1,+1\}^{n}$. However, our analysis reveals that this structure naturally emerges.

\medskip
To verify linearity, the codebook is first translated such that one codeword is mapped to the all-zero vector. Under this translation, binary addition is interpreted as bitwise modulo-2 (or equivalently, multiplication in the $\{-1,+1\}$ domain). We observe that for any pair of codewords $\boldsymbol{x}_i, \boldsymbol{x}_j \in \mathcal{X}$, their sum $\boldsymbol{x}_i \oplus \boldsymbol{x}_j$ (or elementwise product in $\{-1,+1\}$ representation) also belongs to $\mathcal{X}$. This closure property confirms that the learned code is linear, despite no linearity constraint being imposed during training.

\begin{table}[t]
    \centering
    \caption{Hamming distance spectrum of the learned codebook compared to the Hamming$(7,4)$ code.}
    \begin{tabular}{ccccccccc}
        \toprule
        \textbf{Distance} & 0 & 1 & 2 & 3 & 4 & 5 & 6 & 7 \\
        \midrule
        Learned code & 1 & 0 & 0 & 7 & 7 & 0 & 0 & 1 \\
        Hamming$(7,4)$ & 1 & 0 & 0 & 7 & 7 & 0 & 0 & 1 \\
        \bottomrule
    \end{tabular}
    \label{tab:distance_spectrum}
\end{table}

\medskip
Further insight into the code structure is obtained by analyzing its Hamming distance spectrum. The pairwise Hamming distances between all codewords in $\mathcal{X}$ are computed and their distribution is compared with that of the classical Hamming$(7,4)$ code. The learned code exhibits an identical distance spectrum, with minimum Hamming distance $d_{\min} = 3$ and the same number of codewords at each distance, as showin in Table \ref{tab:distance_spectrum}. This implies that the AE not only learns a binary and linear code, but also reproduces the optimal distance properties of the Hamming$(7,4)$ code.

\medskip
These observations demonstrate that gradient-based end-to-end training, combined with the proposed continuous-to-binary learning strategy, is capable of discovering structured and algebraically optimal codes without any prior knowledge of coding theory.

\subsection{Decoding Complexity}
The decoding complexity of the proposed AE-based system differs significantly from that of conventional schemes. The neural decoder consists of a small feedforward network with one hidden layer of size $2^{k}$, requiring two real-valued matrix–vector multiplications and a softmax operation, which is computationally more demanding than binary logic operations used in classical decoders. For the Hamming$(7,4)$ code, brute-force ML decoding, based on computing the Hamming distance between the received word and all $2^{k}=16$ valid codewords, is easily manageable, but its complexity scales exponentially with $k$ and becomes impractical for longer codes. 

\subsection{Discussion}
Although the proposed AE-based framework successfully learns binary codewords that achieve performance comparable to the optimal Hamming$(7,4)$ code, several important observations can be made regarding its stability and scalability. 
First, the use of one-hot message representations causes the input dimension to grow exponentially with $k$ (i.e., $M = 2^{k}$), which limits the scalability of the current architecture to larger block lengths. 
While some recent studies have successfully trained AE-based communication systems for intermediate block lengths (up to $k \approx 100$)~\cite{ninkovic_2025}, these models typically operate in the continuous domain rather than with strictly binary codewords. 
Future work may therefore explore alternative message embeddings or structured input representations that preserve message distinctness while reducing dimensionality.

Moreover, the training process is not entirely deterministic. 
Different training runs may converge to slightly different binary codebooks, some of which exhibit a smaller minimum Hamming distance ($d_{\min} = 2$) than that of the optimal $(7,4)$ code. 
Such cases result in marginally degraded BLER performance. 
This behavior reflects the inherent stochasticity of the optimization process and suggests that convergence toward the optimal discrete structure depends on both initialization and training dynamics. 
Incorporating appropriate regularization or architectural constraints could enhance stability and improve consistency across training instances.

\section{Conclusion}
This paper presented a simplified AE-based framework for learning binary channel codes compatible with practical digital communication systems. By employing a two-stage training strategy, continuous pretraining followed by direct binarization, the proposed approach enables the joint optimization of encoder and decoder networks without relying on gradient approximation techniques. Experimental results for the $(7,4)$ block length show that the proposed method learns a rotated version (coset code) of the optimal Hamming code, naturally reproducing its linear structure and distance spectrum, and achieving the same BLER performance as the Hamming$(7,4)$ code with ML decoding. These findings demonstrate that even simple neural architectures can implicitly learn structured, capacity-approaching binary codes when trained under appropriate constraints. Future work will focus on extending the framework to longer block lengths and exploring alternative message representations to improve scalability.




\end{document}